\documentclass[]{aastex701}
\usepackage{amsmath}
\usepackage{multirow}
\usepackage{hyperref}
\begin{document}

\title{Cloudy With a Chance of Meatballs}

\author{Wolf Cukier}
\affiliation{University of Chicago \\
William Eckhardt Research Center, \\
5640 South Ellis Avenue, Chicago, IL 60637, USA}
\email{tbd}  

\author{Dominic Samra}
\affiliation{University of Chicago \\
William Eckhardt Research Center, \\
5640 South Ellis Avenue, Chicago, IL 60637, USA}
\email{tbd} 

\author{Vighnesh Nagpal}
\affiliation{University of Chicago \\
William Eckhardt Research Center, \\
5640 South Ellis Avenue, Chicago, IL 60637, USA}
\email{tbd} 

\author{Diana Powell}
\affiliation{University of Chicago \\
William Eckhardt Research Center, \\
5640 South Ellis Avenue, Chicago, IL 60637, USA}
\email{tbd} 

\author{Maria Steinrueck}
\affiliation{University of Chicago \\
William Eckhardt Research Center, \\
5640 South Ellis Avenue, Chicago, IL 60637, USA}
\email{tbd} 

\author{Christopher Wirth}
\affiliation{University of Chicago \\
William Eckhardt Research Center, \\
5640 South Ellis Avenue, Chicago, IL 60637, USA}
\email{tbd}

\begin{abstract}

Speculative fiction has long served an inspiration for genuine scientific inquiry.  One notable work that has almost acted in this manner is the the seminal comedic speculative fiction work \textit{Cloudy with a Chance of Meatballs}.  While exoplaneteers reference this work frequently, we have never engaged with the central prediction of this work\ldots until now!  We perform detailed microphysical modeling of meatball clouds, both bare and coated with marinara sauce, and find that while meatball condensation is possible in temperate atmospheres, the meatballs do not quite grow to the sizes predicted by \textit{Cloudy}.   We do find, however, that such meatball condensation, across a large enough planet, would be able to sustain humanity calorically.

\end{abstract}



\section{Introduction} 
Speculative fiction has long served to inspire, predict, and warn of the consequences of future scientific inquiry.  In exoplanets particularly, science fiction's imagination serves a cultural touch point for a large number of our discoveries.  Most notably, \textit{Star Wars} \citep{Lucas1977StarWars} predicted the existence of circumbinary planets long before the discovery of the first one \citep{Wolszczan1992PlanetarySystemMillisecond}---leading to these planets being called ``Tatooine-like.''  Recent work has also suggested that this feedback goes both ways with fictional exoplanets in works of speculative fiction becoming notably less earth like after the discovery of real exoplanets  \citep{Puranen2024ExoplanetsScienceFiction}.


One very important cultural touchstone for studies of exoplanet clouds is the seminal comedic speculative fiction work \textit{Cloudy with a Chance of Meatballs} \citep{Barret1978CloudyChanceMeatballs}, which has since been adapted into a feature length film \textit{Cloudy with a Chance of Meatballs (2009)} \citep{Lord2009CloudyChanceMeatballs}.  Exoplaneteers have long been in scientific and literary conversation with this work, titling their theses, papers, proposals, and talk titles with allusions to this hypothetical precipitation phenomenon.  Examples of these allusive titles include ``Cloudy with a chance of water'' \citep{Wakeford2015CloudyChanceWater}, ``Cloudy with a chance of dust balls'' \citep{Moses2014ExtrasolarPlanetsCloudy}, ``Cloudy with a chance of biosignatures'' \citep{Astrobites2025WeatherForecastHabitable}, ``Partly cloudy with a chance of chondrites'' \citep{Fries2010PartlyCloudyChance}, ``Cloudy with a chance of \ldots ?'' \citep{Loftus2019CloudyChanceRain}, and ``Cloudy with a chance of disequilibrium chemistry'' (GO 11301; PI Splinter).

Despite being in conversation with \textit{Cloudy}, these works all fail to engage with the meat of the text.  Even though investigating a wide range of potential condensates including water, biosignatures and `\ldots ?', none of these works have addressed the obvious question: could meatballs condense in, and then rain out of, exoplanet atmospheres?  We seek to right this catastrophic failure of the exoplanet field by running detailed microphysical models to understand the planetary conditions which give rise to meatball precipitation, as well as characterizing what that precipitation might look like.  In Section \ref{sec:model}, we present the our model of planetary scale meatball condensation; in Section \ref{sec:results} results we present the complete microphysically derived size distributions of meatball clouds and answer important questions such as ``How large of a catchment area is required to sustain human life'' and ``How efficiently does marinara sauce heterogeneously nucleate onto the meatballs''; and finally in Section \ref{sec:conclusion} we will discuss the wide-ranging implications of this work, especially to our understanding of life in the universe.

\section{The Model}\label{sec:model}
Meatball clouds form via complex microphysical processes that depend strongly on planetary properties, notably a planet's thermal structure, culinary composition, and the strength of mixing in the atmosphere. To model condensible meatball clouds in the atmospheres of exoplanets, we use \texttt{CARMApy} (Cukier et al. \textit{in prep.}), a python wrapper of \texttt{ExoCARMA} \citep{Powell2018FormationSilicateTitanium, Gao2018MicrophysicsKClZnS}, a bin-scheme microphysics code that was adapted from the \texttt{CARMA 3.0} \citep{Bardeen2008NumericalSimulationsThreedimensional} update of \texttt{CARMA} \citep{Turco1979OneDimensionalModelDescribing,Toon1988MultidimensionalModelAerosols}.  \texttt{CARMA} treats the microphysical processes of homogeneous nucleation, heterogeneous nucleation, condensational growth, evaporation (searing), and coagulation (meatball fusion), as well as vertical transport of meatball particles due to atmospheric mixing and gravitational settling. 

CARMA resolves the meatball particle size distribution using bin-scheme microphysics. In the bin-scheme approach, the size distribution is discretized into multiple bins according to radius---ranging from sub-micron to potentially bite-sized---and the meatballs in each bin evolve freely and interact with other bins in an Eulerian framework. There is no a priori assumption of the meatball size distribution. 

In our model setup, we assume all atmospheric carbon is in the form of gaseous protein and fat species which initially diffuse from the lower atmosphere until they reach a point in the atmosphere where they become supersaturated and meatball formation occurs via nucleation.  Once these seed particles grow to a large enough radius that heterogeneous nucleation becomes energetically favorable, they become cloud condensation nuclei for the marinara sauce that will coat them. These species will continue to grow through either condensation of ambient fats and proteins, in the case of uncoated meatballs, or gaseous marinara sauce, in the case of coated meatballs.  Meatballs will be mixed upwards through turbulent diffusive forces until gravitational settling becomes dominant and causes the meatballs to fall to hotter parts of the atmosphere where they become overcooked and then vaporized.  We assumed the P-T profile of the planet was that of a $T_{eq}=$ 250 K, log g (cgs) = 4.0 Sonora Bobcat \citep{Morley2024SonoraSubstellarAtmosphere} run with a constant $K_{zz}$ of $10^8$ cm$^2$/s.

\subsection{Laboratory Data}
\begin{figure}
    \centering
    \includegraphics[width=0.7\linewidth]{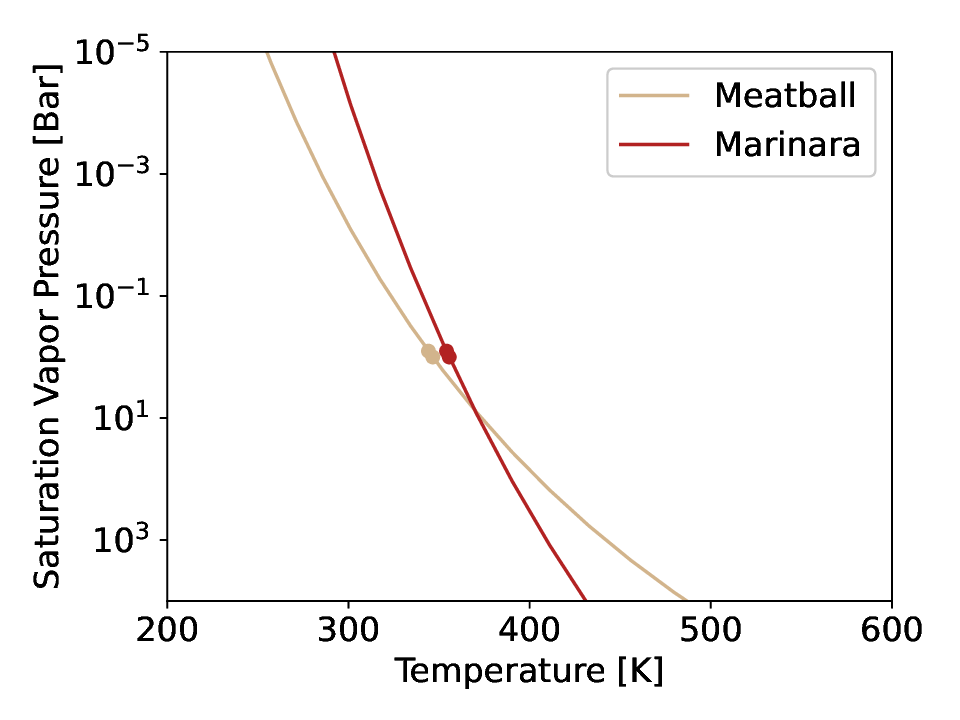}
    \caption{Saturation vapor pressure curves for the meatball and marinara condensates.  Points are the measurements taken in Denver (low-pressure) and Chicago (high-pressure).  The line is the fit to this curve extrapolated into the relevant P-T regimes.}
    \label{fig:svp}
\end{figure}
\texttt{CARMApy} allows for the specification of novel condensates but requires a detailed knowledge of the physical properties of such condensates in order to perform the simulations---including knowledge of properties such as the saturation vapor pressure and surface tension of meatballs which were not readily available in the literature. As our research team was composed entirely of theoretical astrophysicists, we felt highly qualified to take the relevant laboratory data ourselves.  

\begin{figure}
    \centering
    \includegraphics[width=0.7\linewidth]{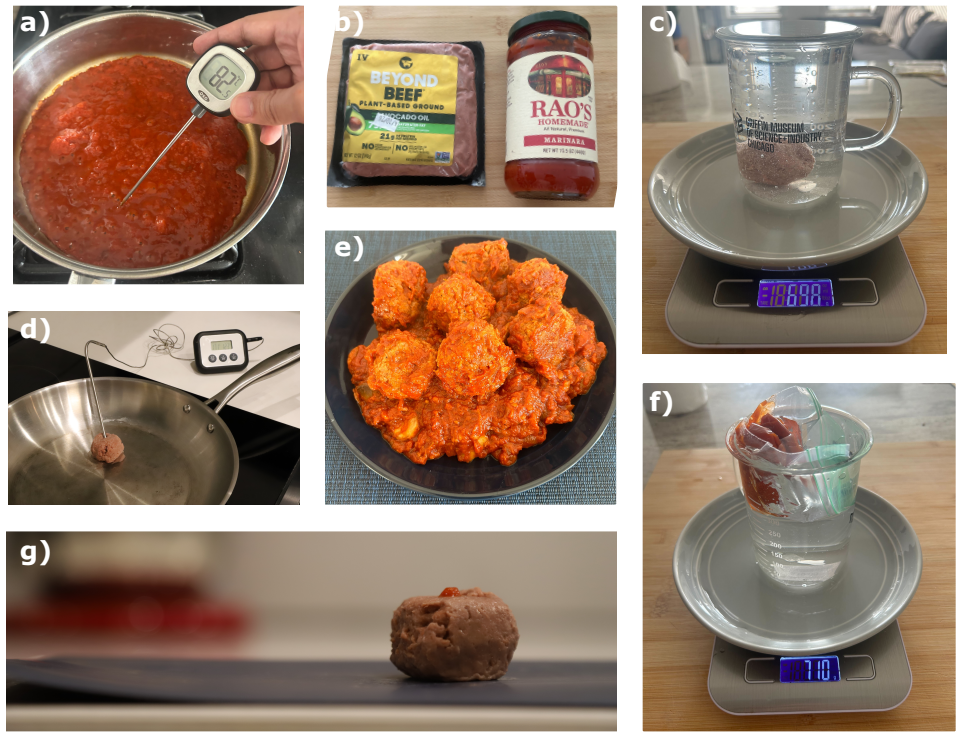}
    \caption{Collage of experimental techniques used to characterise the material properties of meatballs and marinara. a,d) evaporation temperature measurements, b) samples used (not sponsored), c,f) volume measurements, e) served meatballs coated in marinara (core-shell), g) contact angle between meatballs and marinara.}
    \label{fig:experiments}
\end{figure}

To measure the saturation vapor pressure of the meatballs and marinara sauce, we put the condensates in a pan and heated them until they began to vaporize and their temperatures began to plateau (Fig.~\ref{fig:experiments} panels a,d). We note that this procedure, along with all other laboratory procedures involving laboratory data in this work, used vegan meatballs.  Due to our assumption of nucleation from gas-phase fatty acids and proteins, we find that vegan meat based substitutes likely better resemble the qualities of meatballs forming in an atmosphere than beef-based meatballs, especially given the lack of detections flying cows in exoplanet atmospheres to date. This process was repeated both in Denver and Chicago in order to measure the saturation vapor pressure curve at different pressures.  We then fit those points to the following equation to put the saturation vapor pressure in a form that can be used by CARMApy.
\begin{equation}
    \log_{10} \frac{p'}{(10^6 \text{ barye})} = \alpha_0 - \frac{\alpha_1}{T}
\end{equation}
These fits to the saturation vapor pressure curves are presented in Figure~\ref{fig:svp}

{Material densities were experimentally determined through mass and volume measurements. Masses were measured using a state-of-the-art `Torchtree ultra-thin kitchen scale'\footnote{\href{https://www.amazon.co.uk/Electronic-Kitchen-Digital-Stainless-Precision/dp/B08FHRNR1Q?ref_=Oct_d_obs_d_49980854031_5&pd_rd_w=QOiVv&content-id=amzn1.sym.7c9c1e89-3eda-4909-bcfc-af4a63375cec&pf_rd_p=7c9c1e89-3eda-4909-bcfc-af4a63375cec&pf_rd_r=JT9VPAQWCRPYWCS63B7R&pd_rd_wg=AlqDP&pd_rd_r=43a1880e-7991-4bd6-8dab-eb92b89f2aa4&pd_rd_i=B08FHRNR1Q&th=1}{Available at all good online retailers}}. Volume measurements were measured using cutting edge displacement method (Fig.~\ref{fig:experiments} panels c, f), based on the Archimedian principle \citep{ArchimedesofSyracuse530FloatingBodies}.}
{The contact angle between marinara and meatballs was determined by digitally measuring the angle formed by a marinara drop on the meatball surface (Fig.~\ref{fig:experiments} panel g)}.  Surface tensions were determined by assuming that the meatballs are covered in a thin fatty coating with a surface tension similar to well-characterized fatty acids \citep{Kallio2009IntermolecularInteractionsAdhesion} and the surface tension of the marinara sauce is similar to that of water.

\section{Results and Discussion}\label{sec:results}
\begin{figure}
    \centering
    \includegraphics[width=.98\linewidth]{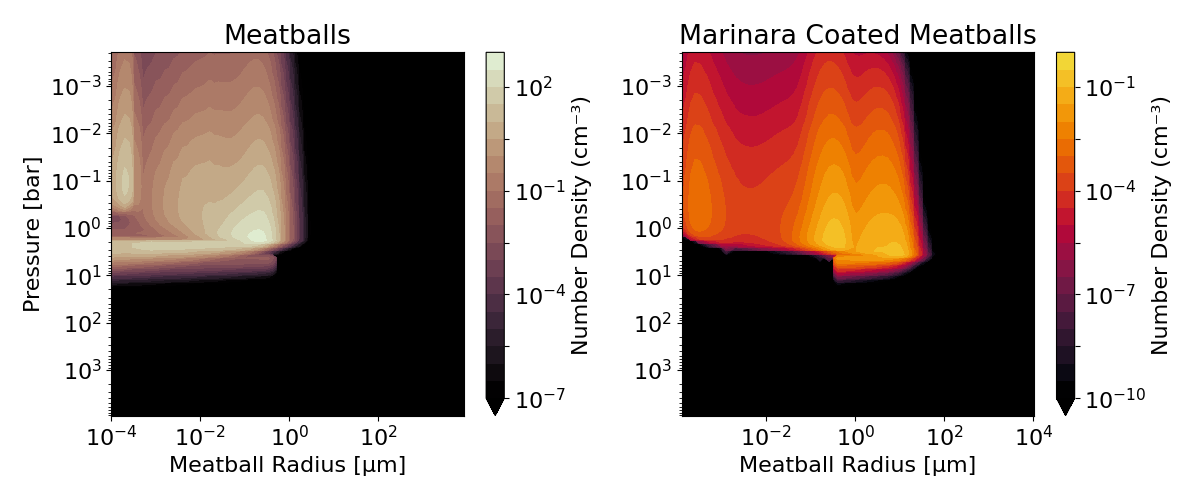}
    \caption{Left: Size distribution for meatball cloud particles as derived from our \texttt{CARMApy} models. Right:  The same but for clouds composed of meatballs coated in marinara sauce.}
    \label{fig:size_dist}
\end{figure}

We present the results of our incredibly rigorous simulations in Figure~\ref{fig:size_dist}.  We find that meatballs can grow to roughly 1 micron in size before they begin to grow so large they rapidly rain out of the atmosphere and evaporate.  The marinara coated meatballs are able to grow larger, up to almost 100 microns.  Both of these sizes fall short, however, of the decimeter to meter sized meatballs predicted by \textit{Cloudy}, meaning such precipitation as predicted therein is exceedingly unlikely, outside of extreme as-of-yet unobserved conditions. 

As these meatballs are forming due to gasses upwelling from the deep atmosphere, these meatballs will be replenished over time.  If we assume that the meatballs refresh themselves on a timescale of $\sim$1 day, we can calculate the catchment area required to sustain a person eating a standard 2000 calorie diet.  The meatball column mass density, both coated and uncoated, is $10^{-8}$ g/cm$^2$ so a person could be sustained by the meatballs located in 667 hectares.  This, correspondingly, means the human population of Earth could be sustained by the meatball condensates from a Jovian sized planet---a food source that is unrivaled by any smaller, terrestrial planets with non-non-arable land.

\section{Conclusions}\label{sec:conclusion}
We have performed rigorous theoretical modeling of the meatball precipitation first proposed by \citet{Barret1978CloudyChanceMeatballs} in their work \textit{Cloudy with a Chance of Meatballs}.  We have presented a-priori size distributions for meatballs, both coated in marinara sauce and bare, that show that while meatball precipitation is possible, it will not reach the sizes predicted by \textit{Cloudy}.  We do note that these meatballs could provide food for a civilization on a planetary scale and thus should be considered as a potential energy source when searching for life in the universe. {The inclusion of spaghetti in future modeling could also allow us to suggest planets in the Spaghetti Habitable Zone (SHZ), allowing us to identify candidate home-worlds for the Flying Spaghetti Monster \citep{Henderson2005OpenLetter}.}

Future work should include modeling of the broad variety of condensates predicted by \textit{Cloudy}. Despite being the titular type of torrent, meatballs are just one of many culinary concoctions condensing in the cult classic. An investigation of brussel sprouts, gorgonzola cheese, pancakes with syrup, and pea soup haze (possibly photochemical in origin), would sate our appetite for discovery. Dedicated GCM (gourmet culinary models) studies could expand on the possibility of tomato tornadoes and extreme salt-and-pepper winds raised by \textit{Cloudy} as well.

\begin{acknowledgments}
All meatballs used in this study were vegan meat substitutes and consumed so as to avoid food waste.  Data for this paper were taken during AASTCS 11, Exoplanet Atmospheres 2026.
\end{acknowledgments}

\bibliography{references, references2}{}
\bibliographystyle{aasjournalv7}
\newpage



\end{document}